\documentstyle[prl,preprint,aps,psfig]{revtex}
\def\mg32{$^{32}$Mg}
\def\hw{$\hbar\omega\,$}
\def\hcm{${H}_{CM}\,$}
\def\ham{${H}\,$}
\begin{document}
\draft
\tighten
\preprint{ }
\title{Shell model Monte Carlo studies of neutron-rich nuclei
in the $1s$-$0d$-$1p$-$0f$ shells}
\author{D.J. Dean$^1$, M.T. Ressell$^{2,4}$,
M. Hjorth-Jensen$^3$, S.E. Koonin$^2$, K. Langanke$^5$, \\ and A.P. Zuker$^6$}
\address{
$^1$Physics Division, Oak Ridge National Laboratory, P.O. Box 2008\\
Oak Ridge, Tennessee 37831 USA and Department of Physics and \\
Astronomy, University of Tennessee, Knoxville, Tennessee, 37996 \\
$^2$W.K. Kellogg Radiation Laboratory, 106-38, California
Institute of Technology \\
Pasadena, California 91125 USA \\
$^3$Nordita, Blegdamsvej 17, DK-2100 Copenhagen \O, Denmark and\\
    Department of Physics, University of Oslo, Norway\\
$^4$Astronomy and Astrophysics Center, University of Chicago \\
5640 S. Ellis Ave., Chicago, IL 60637 \\
$^5$Institute for Physics and Astronomy, Aarhus University, Denmark\\
$^6$IRES, B\^at27, IN2P3-CNRS/Universit\'e Louis
Pasteur BP 28, \\ F-67037 Strasbourg Cedex 2, France}
\date{\today}
\maketitle
\begin{abstract}

We demonstrate the feasibility of realistic Shell-Model Monte Carlo
(SMMC) calculations spanning multiple major shells, using a
realistic interaction whose bad saturation and shell properties
have been corrected by a newly developed general prescription.
Particular attention is paid to the approximate restoration of
translational invariance. The model space consists of the full
$sd$-$pf$ shells. We include in the study some well-known $T$=0
nuclei and several unstable neutron-rich ones around $N = 20,28$.
The results indicate that SMMC can reproduce binding
energies, $B(E2)$ transitions, and other observables with an
interaction that is practically parameter free. Some interesting
insight is gained on the nature of deep correlations.
The validity of previous studies is confirmed.

\end{abstract}

\section{Introduction}

Studies of extremely neutron-rich nuclei have revealed a number of
intriguing new phenomena.  Two sets of these nuclei that have received
particular attention are those with neutron number $N$ in the vicinity
of the $1s0d$ and $0f_{7/2}$ shell closures ($N \approx 20$ and $N
\approx 28$).  Experimental studies of neutron-rich Mg and Na isotopes
indicate the onset of deformation, as well as the modification of the
$N = 20$ shell gap for \mg32 and nearby nuclei \cite{r:motobayashi}.
Inspired by the rich set of phenomena occurring near the $N = 20$
shell closure when $N \gg Z$, attention has been directed to nuclei
near the $N = 28$ (sub)shell closure for a number of S and Ar isotopes
\cite{r:brown1,r:brown2} where similar, but less dramatic, effects
have been seen as well.

In parallel with the experimental efforts, there have been several
theoretical studies seeking to understand and, in some cases, predict
properties of these unstable nuclei.  Both mean-field
\cite{r:werner,r:campi} and shell-model calculations
\cite{r:brown1,r:brown2,r:wbmb,r:poves1,r:fukunishi,r:retamosa,r:caurier}
have been proposed. The latter require a severe truncation to
achieve tractable model spaces, since the successful description of these
nuclei involves active nucleons in both the $sd$- and the $pf$-shells.
The natural basis for the problem is therefore the full $sd$-$pf$
space, which puts it out of reach of exact diagonalization on current
hardware.

Shell-Model Monte Carlo (SMMC) methods
\cite{r:smmc_pr,r:smmc_ar,r:lang} offer an alternative to direct
diagonalization when the bases become very large. Though SMMC provides
limited detailed spectroscopic information, it can predict, with good
accuracy, overall nuclear properties such as masses, total strengths,
strength distributions, and deformation --- precisely those quantities
probed by the recent experiments. It thus seems natural to apply SMMC
methods to these unstable neutron-rich nuclei. Two questions will
arise --- center-of-mass motion and choice of the interaction --- that
are not exactly new, but demand special treatment in very large spaces.

The center-of-mass problem concerns momentum conservation. It
was investigated for the first time by Elliott and Skyrme in
1955~\cite{es}, and a vast literature on the subject has developed,
but as of now, the methods proposed have not managed to reconcile
rigor and applicability. Section II will be devoted to explaining why
this is so and to describe how --- short of ensuring exact momentum
conservation --- it is possible within an SMMC context to assess the
damage and control it in order to perform meaningful calculations.

There has long been a consensus that G-matrices derived from
potentials consistent with $NN$ data \cite{hko95} are the natural shell-model
choice.  Unfortunately, such interactions give results that rapidly
deteriorate as the number of particles increases. Two alternative
cures have been proposed: sets of fitted matrix elements (all the shell-model
work quoted above), or minimal ``monopole'' modifications \cite{acz}.
The latter restricts the fit to far fewer quantities: some average
matrix elements, which are the ones that suffer from the bad
saturation and shell properties of the realistic potentials. Both
approaches have the common shortcoming of needing data to determine
the fitted numbers, but recently a general parametrization of the
monopole field ($H_m$) has become available that could be used to
replace the $G$-matrix centroids for any model space~\cite{dz98}. The
interaction we present in Section III is the first monopole modified
$G$-matrix free of parameters other than the six entering the
independently derived $H_m$.

Section IV contains results for a number of unstable, neutron-rich
nuclei near the $N = 20$ and 28 shell closures and compares them to
experiment and to other truncated shell-model calculations.
Section V is devoted to a discussion of what we have
accomplished and surveys further applications of such
calculations.

\section{SMMC and Center-of-Mass Motion}

By momentum conservation, a many-body wavefunction must factorize as
${\Psi({\bf r})}=\phi(R)\Psi({\bf r}_{rel})$, where $R$ is the center-of-mass
coordinate and ${\bf r}_{rel}$ the relative ones. There are formalisms in
which the latter are constructed explicitly, but they lead to very
hard problems of antisymmetrization. What can be done in a shell-model context
is to work with a basis that ensures that the eigenstates
automatically factorize as requested. This is accomplished
by taking $\phi(R)$ to be a
harmonic oscillator state, which implies that the basis must
produce eigenstates of
\begin{equation}
H_{CM} = {\tilde {P^2}\over{2 A m}} + {1\over{2}} m A
\omega^2 \tilde R^2 - {3\over{2}}\hbar\omega,
\end{equation}
with $\tilde {P}=\sum_{i=1,A} p_i$, and $\tilde {R}=(\sum_{i=1,A} r_i)/A$.

In order to diagonalize this one-body Hamiltonian in the SM basis,
we have to rewrite it using
\begin{equation}
  \label{P2}
(\sum_{i=1}^A p_i)^2=A\sum_{i=1}^A p_i^2-\sum_{i<j}(p_i-p_j)^2,
\end{equation}
along with a similar expression for the coordinates. Then $H_{CM}=h_1+h_2$,
where $h_1$ is a one-body oscillator spectrum, and $h_2$ an oscillator
two-body force. If one considers the matrix element \mbox{$\langle
  n_1l_1n_2l_2|h_2| n_3l_3n_4l_4\rangle$}, it is quite easy to
convince oneself --- using a general property of the Talmi-Moshinsky
transformation and the oscillator form of $h_2$ --- that
\mbox{$2n_1+l_1+2n_2+l_2=2n_3+l_3+2n_4+l_4$}. In other words, $H_{CM}$
conserves the number of oscillator quanta. This implies that if a basis
contains {\em all} states of (or up to) $n$\hw excitations,
diagonalizing a translationally invariant Hamiltonian would ensure the
factorization of the center-of-mass wavefunction. To separate the
wavefunctions
with 0\hw center-of-mass quanta it would be sufficient to do the calculations
with
\begin{equation}
{\tilde H} = H + \beta_{CM} H_{CM}
\label{GL}
\end{equation}
choosing a large $\beta_{CM}$ (not to be confused with the SMMC
inverse temperature). Thus, the procedure to deal with the center-of-mass
problem is conceptually straightforward. Practically, things are not
so simple.  In \mg32, for example, the $sd$-$pf$ basis will contain
states having between 0 and 16\hw quanta; however, it is very far from
containing them all, and it does not even contain all those of 1 \hw.
Then, and this point is crucial, the restriction of
$H_{CM}$ to the basis {\bf is no longer} $H_{CM}$. As a consequence,
the prescription in Eq.~(\ref{GL}) is no longer a prescription to
remove unwanted center-of-mass excitations, but a prescription to remove
something else.  Still, whatever the restricted $H_{CM}$ is in the
model space, it is the operator most closely connected with the true
one. Hence, rather than removing unwanted excitations, which is now
impossible in general, we may try to assess and control the damage by
using Eq.~(\ref{GL}) to construct a set of states $|\beta_{CM}
\rangle$ and see how $\langle \beta_{CM}|H|\beta_{CM} \rangle$
behaves. Since the problem is variational, the best we can do is
choose a $\beta_{CM}$ that minimizes the energy.

Before proceeding, it is worth going quickly through the history of
the subject, under the light of these very elementary considerations
which are often ignored, thereby creating unnecessary confusion.  The
pioneers of the subject were Elliott and Skyrme~\cite{es} who treated
a simple case, the 1 \hw $JT^{\pi}$=$1^-0$ excitations on a $sp$ shell
core, showing that one of them ($\sqrt{5\over 6}p^{-1}d-\sqrt{1\over
  6}p^{-1}s$) was simply the 1 \hw center-of-mass state. Other early important
contributions are~\cite{gs57,r:baranger}.  The first
cross-shell calculation in a full space appeared
in 1968~\cite{zbm}: $(p_{1/2}s_{1/2}d_{5/2})^n$, which
successfully accounted for the
spectra in the region around $^{16}$O.
The $(p_{1/2}^{1}s_{1/2})J^{\pi}T=1^-0$ state contained a spurious
component of 5.556\% of the Elliott-Skyrme state.  Nonetheless, Gloeckner and
Lawson~\cite{r:gloeckner} decided to apply Eq.~(\ref{GL}) with an
arbitrarily large $\beta_{CM}$ to eliminate the spurious components; by not
realizing that $H_{CM}$ restricted to that small space generated very
little center-of-mass excitations and many genuine ones, they managed to
eliminate the latter rather than the former. In spite of the criticism
that ensued~\cite{r:mcgrory,r:whitehead} showing that the procedure
could not possibly make sense (except in complete spaces), no
formally satisfactory arguments were advanced to replace it.
Equation~(\ref{GL}) remained a guide on where to begin to
minimize the center-of-mass nuisance, and it is indeed the basis of our
variational suggestion. New projection techniques have been developed
\cite{r:rath}, but they rely on explicit construction of the spurious
states and they are not applicable in SMMC.

The recipe advocated by Whitehead {\it et al.}  \cite{r:whitehead}
seems quite compatible with the constrained variation sketched above,
and we have adopted it, since it proves sufficient to optimize the
solutions. The idea is to add $\beta_{CM}$\hcm to \ham,  but with
$\beta_{CM}$ remaining fairly small.  We have found that $\beta_{CM} =
1$ works reasonably well.  This value will push spurious components up
in energy by $\hbar\omega = 45 A^{-1/3} - 25 A^{-2/3}$ MeV $\simeq 14$
MeV while leaving the desired components relatively unscathed.  A
smaller value of $\beta_{CM}$ leaves the spurious configurations at
low enough energies that they are included in the Monte Carlo
sampling, while larger values of $\beta_{CM}$ ($>3$) begin to remove
the entire $pf$ shell from the calculation and artificially truncate
the space. Technically, \hcm suffers from the sign problem, and we
have to say a word about it.

SMMC methods reduce the imaginary-time, many-body evolution operator to
a coherent superposition of one-body evolutions in fluctuating
one-body fields.  This reduction is achieved via a
Hubbard-Stratonovich transformation and the resulting path integral is
evaluated stochastically.  SMMC methods have been applied to numerous
full-basis 0\hw studies.  The primary difficulty in these applications
arises from a sign problem due to the repulsive part of effective
nucleon-nucleon interactions.  A practical solution to this sign
problem was obtained by considering a set of Hamiltonians close to the
desired realistic Hamiltonian ($H$) and extrapolating to the
realistic case \cite{r:sign}.  This technique has been validated in
numerous studies that show the SMMC approach to be a viable and
productive avenue to study extremely large many-body problems
\cite{r:smmc_pr,r:smmc_ar,r:lang}.

The original sign problem for realistic interactions was solved by
breaking the two-body interaction into ``good'' (without a sign
problem) and ``bad'' (with a sign problem) parts: $ H = H_{good} +
H_{bad}$.  The bad part is then multiplied by a parameter, $g$, with
values typically lying in the range $-1 \le g \le 0$.  The Hamiltonian
$ H = f(g) H_{good} + g H_{bad}$ has no sign problem for $g$ in this
range.  The function $f(g)$ is
used to help in extrapolations. It is constructed such that $f(g=1)=1$, and
takes the form $[1-(1-g)/\chi]$, with $\chi=4$.
The SMMC observables are evaluated for a number of different
negative $g$-values and the true observables are obtained by
extrapolation to $g = 1$.  If we fix the sign problem in the same
manner as above for \hcm we are no longer dealing with a Hamiltonian
that pushes {\it all} spurious components to higher energies --- some
components might even be lowered for $g < 0$. We will see shortly that
this is not a real problem.

We typically choose a minimal extrapolation (linear, quadratic, etc.)
in the extrapolation parameter that gives a $\chi^2$ per datum of
$\simeq 1$. In much of our work most quantities extrapolate either
linearly or quadratically.  We measure the center-of-mass contamination by
calculating the expectation value of $H_{CM}$.  In Fig.~1a, we show
the value of $\langle H_{CM} \rangle$ in \mg32 for several different
values of $\beta_{CM}$.\footnote{All calculations presented here
were performed in the zero temperature formalism \cite{r:lang}
using a cooling parameter of $1/\beta =
0.5$~MeV with $\Delta\beta = 1/32$~MeV$^{-1}$.  These values have
been shown to be sufficient to isolate the ground state for
even-even nuclei.  For all data presented here 4096 samples were
taken at each value of the extrapolation parameter, $g$.}  It is
apparent that $\langle$\hcm$\rangle$ decreases as $\beta_{CM}$
increases.  We also find that near $\beta_{CM} = 1$, $\langle H_{CM}
\rangle \ll 2\hbar\omega \simeq 28$ MeV showing that the center-of-mass
contamination
is minimal.  Note that at approximately $\beta_{CM}=1.5$ the average
of the two different techniques of extrapolation presented in Fig.~1a
give $\langle H_{CM} \rangle \simeq 0$~MeV, and the calculations could
be fine tuned for each nucleus to obtain this value.

Figure 1a contains two different data sets corresponding to
two different methods of extrapolating $\langle  H_{CM}\rangle $
to the  physical case ($g = 1$).
The solid circles show the results of a
simple linear extrapolation where for this
observable $\chi^2$ per datum is approximately 1.
It has been established \cite{r:smmc_pr} that
$\langle  H \rangle$ obeys a variational principle such
that the extrapolating curve must have a minimum
(slope = 0) at the physical value ($g=1$).  As we
 sample values of
the quantity ${\tilde H}$,
it is perhaps reasonable to extrapolate $\langle  H_{CM} \rangle$
using this constraint as well (if $ {\tilde H}$ were truly separable,
this would be an exact procedure).
A cubic extrapolation embodying this constraint corresponds to the
open circles in Fig.~1a.

We may further evaluate our extrapolation procedures by comparing SMMC
and the standard shell-model results in $^{22}$Mg.  Shown in
Fig.~2a is a detailed comparison for the expectation of the energy
$\langle H\rangle$, and in Fig.~2b a comparison for $\langle
H_{CM}\rangle$.  The standard shell-model results were obtained using the
code ANTOINE
\cite{r:antoine}.  The SMMC results in Fig.~2a employ a constrained
fit, such that $d\langle H\rangle/dg \mid_{g=1}=0$. The slight
deviation from the standard shell model at $g=-0.6,-0.8,-1.0$ is due to
increasing
interaction matrix elements (with $g$), while $d\beta$, the imaginary
time step, is kept fixed. This deviation is also seen in Fig.~2b. Note
that in Fig.~2b neither the constrained fit nor the linear fit (both
with $\chi^2$ per datum $\simeq 1$) give a precise description of the
standard shell-model results at $g=1$. An average
of the two ways of extrapolation as
indicated by the solid line on Fig.~2b, apparently gives the more
precise result, and we shall do this for other $H_{CM}$ values quoted
throughout this paper. The error bar for such an averaged result is given
by adding in quadrature the individual errors of both extrapolations.

In Fig.~1b we show the evolution of the total $B(E2)$  and in
Fig.~1c we show the occupation of the $sd$-shell and the
$f_{7/2}$-shell as a function of $\beta_{CM}$.  Note that the
occupation of the $f_{7/2}$ orbit decreases as $\beta_{CM}$ increases.
This is due to a combination of the removal of actual center-of-mass
excitations
and the ``pushing up'' in energy of the real states.  The $B(E2)$
decreases slowly with $\beta_{CM}$, although the uncertainties are
consistent with a constant. However, the decrease, particularly at
$\beta_{CM}=3$, is likely to be real since we are working in an
incomplete $n$\hw model space.  At extremely large values of
$\beta_{CM}$ we would remove the $pf$ shell from the calculation and
return to the pure $sd$-shell result, which is substantially smaller
than the result shown here.  The slow evolution of the $B(E2)$ with
$\beta_{CM}$ does open the intriguing possibility of studying
$B(E2)$s with an interaction that has no sign problem (e.g. Pairing +
Quadrupole) and no center-of-mass correction with the hope of obtaining
reasonable results.

Somewhere between $\beta_{CM}= 3$ and 5, $\beta_{CM} H_{CM}$ begins to
change so strongly as a function of $g$ that our extrapolations become
unreliable and we can extract no useful information.  By
$\beta_{CM}=5$, the extrapolated values become completely
unreasonable, and numerical noise completely swamps the calculation.
We thus conclude that a safe value for a generic study is
$\beta_{CM}=1$, although for a given nucleus this value may be fine
tuned to nearly eliminate all center-of-mass contamination from the
statistical observables. This may be done in future studies.

\section{The Effective Interaction}

Numerous shell-model studies have been carried out in truncated model
spaces for neutron-rich nuclei near $N = 20$
\cite{r:wbmb,r:fukunishi,r:poves1} and $N = 28$
\cite{r:brown1,r:brown2,r:retamosa}.  The number of $sd$-$pf$ shell
effective interactions used almost exceeds the number of papers, but
there are similarities between them. A common feature is Wildenthal's
USD interaction \cite{r:wildenthal} to describe the pure $sd$ shell
part of the problem.  All also use some corrected version of the
original Kuo-Brown (KB) $G$-matrix interaction \cite{KB} to describe
nucleons in the $pf$ shell.  The cross-shell interaction is handled in
one of two different ways: matrix elements are generated via a
$G$-matrix or via the Millener-Kurath potential.  As is common in
this type of calculation, selected two-body matrix elements and single-particle
energies have been adjusted to obtain agreement with
experiment. As these interactions have been produced for use in highly
truncated spaces (usually with only $2p2h$ neutron excitations to the
$pf$-shell), they are not suitable for use in the full space. We found
that they generally scatter too many particles from the $sd$ to the
$pf$ shell, and that the $B(E2)$ values cannot be consistently
calculated. We are not saying that the interactions are wrong, but
that we did not succeed in adapting them to the full space.
Perhaps it can be
done, but it would be of limited interest; the years of experience
these forces embody cannot be transposed to other spaces, as the
$pf$-$sdg$ shells for instance, where detailed fits are unthinkable.
Therefore, we derived a new effective interaction for the region: a
monopole corrected renormalized $G$-matrix, derived from a modern
potential.

As noted in Section I, if $G$-matrices have not been widely used, it
is because they were thought to be so flawed as to serve at best as
input parameters to overall fits, as in the
case of the famous USD interaction
\cite{r:wildenthal}. However, it had been pointed out twenty years ago
\cite{pasquini} that practically all the problems of the KB
interaction amounted to the failure to produce the $N,Z=28$ closure,
and could be corrected by changing at most four centroids of the
interaction. A
perturbative treatment in the beginning of the $pf$ shell using these
modifications (the KB3 interaction) gave good results \cite{PZ81a},
and when the ANTOINE code became available \cite{r:antoine}, the
results became truly excellent (\cite{4749,phila} and references
therein). In the meantime, it was confirmed in other regions that
the only trouble with the $G$-matrices resided in their centroids,
i.e., in the bad saturation and shell formation properties of the
realistic potentials~\cite{acz}. The rest of the interaction was
excellent, and strongly dominated by collective terms (pairing and
quadrupole mainly) \cite{mdz}. We say interaction and not
interactions, because all the realistic ones produce very similar good
multipole matrix elements, and similar monopole failures. The
outstanding problem is to replace case by case modifications by a
general specification of $H_m$, the monopole field, that yields
all the centroids once and for all. In subsection B we shall describe
the proposed solution \cite{dz98} we have adopted.

There is great advantage in the SMMC context to adopt the schematic
collective multipole Hamiltonian ($H_M$) of \cite{mdz}, because its
main terms have good signs, thereby eliminating extrapolation
uncertainties; however, it may be a premature step.
For one thing, it has not been
established yet that in the light nuclei the collective contribution
is sufficient to give high-quality results, a project better left to
exact diagonalizations where fine details may be better probed.
Furthermore,
the renormalization treatment in \cite{mdz} is somewhat crude. A
more complete treatment might yield significant differences. This
could be true even though
potentials consistent with the $NN$ data yield very
similar collective contributions and are
therefore reasonably well fitted even by older potentials.  Finally,
even if it were true that realistic interactions are interchangeable,
and that a crude treatment of renormalization was adequate, there
would certainly be no objection to using the best forces and the most
sophisticated renormalizations available. In practice this is what we do
here.

\subsection{The renormalized $G$-matrix}

In order to obtain a microscopic effective shell-model interaction
which spans both the $1s0d$ and the $0f1p$ shells, our many-body
scheme starts with a free nucleon-nucleon interaction $V$ which is
appropriate for nuclear physics at low and intermediate energies.  At
present there are several potentials available. The most recent
versions of Machleidt and co-workers \cite{cdbonn}, the Nimjegen group
\cite{nim}, and the Argonne group \cite{v18} have a $\chi^2$ per datum
close to $1$ with respect to the Nijmegen database
\cite{nimbase}.  The potential model of Ref.\ \cite{cdbonn} is an
extension of the one-boson-exchange models of the Bonn group
\cite{mac89}, where mesons like $\pi$, $\rho$, $\eta$, $\delta$,
$\omega$, and the fictitious $\sigma$ meson are included. In the
charge-dependent version of Ref.\ \cite{cdbonn}, the first five mesons
have the same set of parameters for all partial waves, whereas the
parameters of the $\sigma$ meson are allowed to vary. The recent
Argonne potential \cite{v18} is also a charge-dependent version of the
Argonne $V14$ \cite{v14} potential. The Argonne potential models are
local potentials in coordinate space and include a $\pi$-exchange plus
parametrizations of the short-range and intermediate-range part of the
potential. The Nimjegen group \cite{nim} has constructed potentials
based on meson exchange and models parametrized in similars ways as
the Argonne potentials.  Another important difference between, e.g.,
the Bonn potentials and the Argonne and Nimjegen potentials is the
strength of the much debated tensor force \cite{bm95}. Typically, the
Bonn potentials have a smaller $D$-state admixture in the deuteron
wave function than the Argonne and Nimjegen potentials, as well as
other potential models. A smaller (larger) $D$-state admixture in the
ground state of the deuteron means that the tensor force is weaker
(stronger).  The strength of the tensor force has important
consequences in calculations of the binding energy for both finite
nuclei and infinite nuclear matter (see, e.g., the discussion in Ref.\
\cite{hko95}).  A potential model with a weak tensor force tends to
yield more attraction in a nuclear system than a potential with a
strong tensor force; however, all these modern nucleon-nucleon
interactions yield very similar excitation spectra. Moreover, in
calculations of Feynman-Goldstone diagrams in perturbation theory, a
potential with a weak tensor force tends to suppress certain
intermediate states of long-range character, like particle-hole
excitations \cite{sommer81}.  In this paper, we choose to work with
the charge-dependent version of the Bonn potential models, as found in
Ref.\ \cite{cdbonn}.

The next step
in our many-body scheme is to handle
the fact that the repulsive core of the nucleon-nucleon potential $V$
is unsuitable for perturbative approaches. This problem is overcome
by introducing the reaction matrix $G$ given by the solution of the
Bethe-Goldstone equation
\begin{equation}
    G=V+V\frac{Q}{\omega - H_0}G,
\end{equation}
where $\omega$ is the unperturbed energy of the interacting nucleons,
and $H_0$ is the unperturbed Hamiltonian.
The projection operator $Q$, commonly referred to
as the Pauli operator, prevents the
interacting nucleons from scattering into states occupied by other nucleons.
In this work, we solve the Bethe-Goldstone equation for several
starting
energies $\Omega$, by way of the so-called double-partitioning scheme
discussed in  Ref.\ \cite{hko95}.
For the closed-shell core in the $G$-matrix calculation
we choose $^{16}$O and employ a harmonic-oscillator basis for the
single-particle
wave functions, with an oscillator energy $\hbar\Omega$ given
by
$\hbar\Omega = 45A^{-1/3} - 25A^{-2/3}=13.9 $ MeV,
$A=16$ being the mass
number.

Finally, we briefly sketch how to calculate an effective
two-body interaction for the chosen model space
in terms of the $G$-matrix.  Since the $G$-matrix represents just
the summation to all orders of ladder diagrams with particle-particle
diagrams, there are obviously other terms which need to be included
in an effective interaction. Long-range effects represented by
core-polarizations terms are also needed.
The first step then is to define the so-called $\hat{Q}$-box given by
\begin{eqnarray}
&&   P\hat{Q}P=PGP+   \label{eq:qbox}\\
&&   P\left(G\frac{Q}{\omega-H_{0}}G +G
   \frac{Q}{\omega-H_{0}}G \frac{Q}{\omega-H_{0}}G +\dots\right)P.
   \nonumber
\end{eqnarray}
The $\hat{Q}$-box is made up of non-folded diagrams which are irreducible
and valence linked.
We can then obtain an effective interaction
$H_{\rm eff}=\tilde{H}_0+V_{\rm eff}$ in
terms of the $\hat{Q}$-box
with \cite{hko95}
\begin{equation}
    V_{\rm eff}(n)=\hat{Q}+{\displaystyle\sum_{m=1}^{\infty}}
    \frac{1}{m!}\frac{d^m\hat{Q}}{d\omega^m}\left\{
    V_{\rm eff}^{(n-1)}\right\}^m,
    \label{eq:fd}
\end{equation}
where $(n)$ and $(n-1)$ refer to the effective interaction after $n$
and $n-1$ iterations. The zeroth iteration is represented by just the
$\hat{Q}$-box.  Observe also that the effective interaction $V_{\rm
  eff}(n)$ is evaluated at a given model space energy $\omega$, as is
the case for the $G$-matrix as well. Here we choose $\omega =-20$ MeV.
The final interaction is obtained after folding results in eigenvalues
which depend rather weakly on the chosen starting energy (see, e.g.,
Ref.\ \cite{converge93} for a discussion).  All non-folded diagrams
through second-order in the interaction $G$ are included.  For further
details, see Ref.\ \cite{hko95}.  Finally, the reader should note that
when one defines an effective interaction for several shells, the
effective interaction may be strongly non-hermitian.  This
non-hermiticity should arise already at the level of the $G$-matrix.
However, since the $G$-matrix is calculated at a fixed starting energy
for both incoming and outgoing states, it is by construction
hermitian. Since we are calculating an effective interaction at a
fixed starting energy, the individual diagrams entering the definition
of the $\hat{Q}$-box are thereby also made hermitian. The
non-hermiticity which stems from folded diagrams is made explicitly
hermitian through the approach of Suzuki {\em et al.\ } in Ref.\
\cite{kenji}.

\subsection{The monopole field}

As results concerning the monopole field are scattered through many
papers~\cite{acz,mdz,zuk94,jdz95}, the most relevant of which is not
yet published~\cite{dz98}, this subsection offers a compact
presentation of the main ideas.

The centroids we have often mentioned are --- in a neutron proton (np)
representation --- the average matrix elements
\begin{equation}
 V^{xx'}_{rs}=\frac{\sum_J (2J+1)V^{Jxx'}_{rsrs}}{ \sum_J
  (2J+1)},
\end{equation}
where $xx'$ stands for neutrons or protons in orbits $rs$ {\em
  respectively}. Technically, the monopole field $H_m$ is that part of
the interaction containing all the quadratic two-body forms in scalar
products of fermion operators $a^+_{rx}\cdot a_{sx'}$ (same parity for
$r$ and $s$). The clean extraction of these forms from the total $H$
  (i.e., the separation $H=H_m+H_M$, $M$ for multipole)
is not altogether trivial \cite{zuk94}. It is conceptually important
because it makes $H_m$ closed under unitary transformations of the
$a^+,a$ operators, and therefore closed
under spherical Hartree-Fock
variation. The expectation values we may want to vary are those of the
$H_m^d$, the diagonal part of $H_m$ in a given basis.  Calling
$m_{rx}$ the $x$-number operator for orbit $r$, we obtain
\begin{equation}
\label{hm}
H_m^d= \sum_{rx,sx'}
V^{xx'}_{rs}m_{rx}(m_{sx'}-{\delta}_{rs}{\delta}_{xx'}),
\end{equation}
a standard result (it is the extraction of the non diagonal terms that
is more complicated). The expectation value of $H_m^d$ for any state is the
average energy of the configuration to which it belongs (a
configuration is a set of states with fixed $m_{rx}$ for each orbit).
In particular, $H_m^d$ reproduces the exact energy of closed shells
($cs$) and single-particle (or hole) states built on them ($(cs)\pm
1$), since for this set ($cs\pm 1$) each configuration contains a
single member. Consequently, it is uncontaminated by direct
configuration mixing. As an example, in $^{56}$Ni, the two-body (no
Coulomb) contribution to the binding energy in the $pf$ shell is approximately
73 MeV, and configuration mixing (i.e., $H_M$) is responsible for
only 5 MeV; the rest is monopole. If we compare to the {\em total}
binding of 484 MeV, it is clear that the monopole part becomes
overwhelming, even allowing for substantial cross shell mixing (which,
incidentally is included in the present calculations).

Therefore, $H_m^d$ is responsible entirely for the bulk $O(A)$ and surface
energies $O(A^{2/3})$, and for a very large part of the shell effects
[$O(A^{1/3})$, i.e., the 73 MeV]. There can be little doubt this is
where the trouble comes in the realistic potentials.

In a nutshell, the idea in \cite{dz98} is to fit $H_m^d$ to the
$(cs)\pm 1$ set, the single-particle and single-hole spectra around
doubly magic nuclei. It is assumed that the bulk and surface terms can be
separated, and by cancelling the kinetic energy
\mbox{$K=\hbar\omega/2\sum_p {(p+3/2)m_p}$}, $m_p$ is the number of
particles in harmonic oscillator (HO) shell $p$, against  the
collective monopole term \cite{mdz,jdz95}, the leading term in $H_m$.
Defining \mbox{$W=\hbar\omega(\sum_p{(m_p/\sqrt{D_p})^2}/4$}, one obtains an
expression of order $O(A^{1/3})$ that has strong shell effects
producing the HO closures. To this one adds $l\cdot s$ and $l\cdot l$
one-body terms that produce the observed splittings around HO
closures. The filling order is now established, and as the largest
orbit---which comes lowest---is full, it alters significantly the
splitting of its neighbors (e.g.,
the spectrum of $^{57}$Ni is totally
different from that of $^{41,49}$Ca). This is taken care of by
strictly two-body terms. With a total of six parameters (two for the
$W-4K+l\cdot s+l\cdot l$ part, and four for the two-body
contributions), the fit yields an $rms$
deviation of 220 keV for 90 data points.

All terms have a common scaling in \hw=$40/\rho$, obtained using a
very accurate fit to the radii $\langle r^2\rangle =0.9\rho A^{1/3}$,
where
\begin{equation}
\label{rho}
\rho=(A^{1/3}(1-(2T/A)^2)e^{(3.5/A)}.
\end{equation}
Note that due to this scaling it is possible to use the
same functional form from $A=5$ to $A=209$.

Figure \ref{f:shell} shows the mechanism of shell formation for nuclei
with $T=4$. There is an overall {\em unbinding} drift of $O(A^{1/3})$,
with pronounced HO closures due to $W-4K$ at $(N,Z)=$ (16,8), (20,12),
and (28,20). The addition of the $l\cdot s+l\cdot l$ terms {\em
  practically
destroys} the closures except for the first ($^{24}$O),
and creates a fictitious one at $^{40}$S. It is only through the two-body
terms that closure effects
reappear, but now the magic numbers are 6 ($^{20}$C), 14 ($^{20}$C,
$^{36}$Si), and 28 ($^{48}$Ca,
$^{64}$Ni). Note that the shell effect
in $^{32}$Mg is minuscule. The same is true for
$^{30}$Ne among the $T=5$ nuclei.
Among the four two-body terms in $H_m$, there is one that is
overwhelmingly responsible for the new (EI, for intruder, extruder)
magic numbers.  It produces an overall (T=1, mainly) repulsion between
the largest (extruded) orbit of the shell and the others. The extruder
becomes the intruder in the shell below. {\em This is the term that is
missing in the realistic interactions}.
The problems in the excitation spectra of $^{47}$Ca, $^{48}$Ca, and
$^{49}$Ca~\cite{pasquini,hko95} disappear if the realistic centroids
are replaced by those---even more realistic, apparently---of $H_m$.

To close this subsection we give some useful formulas to relate the
$np$ and isospin ($mT$) representations. We have
\begin{eqnarray}
&&{H}_{mT}^d =K+\sum_{r\leq s}{1 \over (1+\delta_{rs})}
\bigl[a_{rs}\,m_r(m_s-\delta_{rs})+\nonumber\\
&&             +b_{rs}(T_r\cdot  T_s-{3\over 4}m\delta_{rs})\bigr],
\label{hmt}
\end{eqnarray}
which reproduces the average energies of configurations at fixed $m_rT_r$.

Calling $D_r=2j_r+1$ the degeneracy of orbit $r$, we rewrite the
relevant centroids incorporating explicitly the Pauli restrictions
\begin{eqnarray}
V_{rx,sx'}&=&{\sum_J V_{rsrs}^{Jxx'}(2J+1)
  ( 1-(-)^J\delta_{rs}\delta_{xx'})\over
  D_r(D_s-\delta_{rs}\delta_{xx'})}\nonumber\\
V_{rs}^T&=&{\sum_J V_{rsrs}^{JT}(2J+1)
( 1-(-)^{J+T}\delta_{rs})\over D_r(D_s+\delta_{rs}(-)^T)}
\label{vnp}
\end{eqnarray}
\begin{equation}
a_{rs}={1\over 4}(3V^1_{rs}+V^0_{rs}), \quad
b_{rs}=V^1_{rs}-V^0_{rs}.
\label{ab}
\end{equation}

In the np scheme each orbit $r$ goes into two $rx$ and $rx'$
and the centroids can be obtained through
 $x\ne x'$)
\begin{eqnarray}
V_{rx,sx'}&=&{1\over 2}\left\lbrack V_{rs}^1\left(1-{\delta_{rs}\over D_r}
\right)+ V_{rs}^0\left(1+{\delta_{rs}\over D_r}
\right)\right\rbrack\nonumber\\
V_{rx,sx}&=&V_{rs}^1.
\label{npmt}
\end{eqnarray}

\subsection{The monopole terms in the calculations}

The calculations used the preliminary version of $H_m$ \cite{dz98},
which for the purposes of this study should make little difference.
All we have said above is valid for both the old and the new version
except for details. Only one of them is worth mentioning here, and it
concerns the single-particle energies shown in Table~\ref{t:sp}. It
is seen that the old and new values are quite close to those adopted
in the calculations, though the old set puts the $s_{1/2}$ and
$f_{5/2}$ orbits higher. This reflects the awkward behavior of the
$l\cdot l$ part of $H_m$ that changes sign at the $p=3$ shell. This
problem was treated artificially in the old version through a
single-particle mechanism that was discarded in the calculations,
mainly because keeping it would have demanded a readjustment of the
interaction for each nucleus --- an unwanted complication in a
feasibility study such as this one. As a consequence, we expect the
$s_{1/2}$ orbit to be overbound with respect to its $sd$ partners in
the upper part of the shell. In the new version the mechanism becomes
two-body and should do much better.

There has been much discussion about the choice of the cross-shell
gap, i.e., the distance between the $d_{3/2}$ and $f_{7/2}$ orbits,
which plays a crucial role in all truncated calculations. It could be
thought from Table~\ref{t:sp} that it is rather small. But this is an
illusion since $H_m$ will make it evolve. In $^{29}$Si it will
increase to 4.5 MeV ($\approx 500$ keV above experiment), which grows
up to 5.2 MeV in $^{40}$Ca, now too small with respect to the binding
energy (BE) difference 2BE($^{40}$Ca)-BE($^{41}$Ca)-BE($^{39}$Ca)= 7.2
MeV.  The only way to decide whether these positionings are correct is
through calculations such as the present ones. We return to this issue
in Section IV.

The analysis of binding energies is a delicate exercise because
external parameters have to be introduced. The philosophy behind $H_m$
is to make all calculations {\em coreless}.  Because of the \hw
propagation (which should be extended to $H_M$), nuclei readjust their
sizes and energies as $N$ and $Z$ change. If the bulk terms are added,
there is, in principle, no need to fit anything, and the calculated
energies are absolute --- not referred to any core. In the present
calculations the interaction was kept fixed, and the way to proceed is
the traditional one, by referring all energies to the core of $^{16}$O.
First, we estimate Coulomb effects using $V_c=0.717
Z(Z-1)(A^{-1/3}-A^{-1})$, and then fit
\begin{equation}
H_{\rm corr}=\varepsilon m + a\frac{m(m-1)}{2} + b\frac{(T(T+1)-3m/4)}{2}\;.
\end{equation}
It is generally assumed that $\varepsilon$ should be close to the
single-particle energy of $^{17}$O (-4.14 MeV), and that the
quadratic terms are the average $H_{mT}$ over the space (from
Eq.~(\ref{hmt})). However, these assumptions do not apply here. The
contribution to $b$ from $H_{mT}$ is relatively small. The symmetry
energy must be counted as one of the bulk terms, and the best we can
do is to take it from fits to the binding energies, which yield
consistently similar numbers. From \cite{jdz95} we adopt the form
\mbox{${\cal S}=22[4T(T+1)](1-1.82/A^{1/3})/A$}, where the main term
has been reduced by the approximately 6 MeV coming from $H_m$. We
cannot change these parameters; we can only check that the
fit to $H_{\rm corr}$ yields a $b$
consistent with them. But there is subtlety: the
isospin term vanishes at $m=1$ because it is taken to be two-body,
while ${\cal S}$ gives a substantial 1.15 MeV contribution at
$^{17}$O.  Therefore, to use the form of $H_{\rm corr}$, $\varepsilon$
must be ${\cal S}$-corrected (in the same sense that we
Coulomb-correct) to\  -4.14-1.15=-5.29 MeV.  For $b$ we must take some
average ${\cal S}$, which we choose to be the value at $A=40$, i.e.,
$b=2.34$. Finally for $a$ we must expect a small value, since it
should come entirely from $H_m$. The fit yields
(in MeV)  $\varepsilon=-5.34$, $a=-0.319$, $b=1.99$, and a
$\chi^2$ per degree of freedom of $3.12$. While $\varepsilon$ and $b$
are very comfortably close to our expectations, $a$ is much too
large. But that is not a problem: the program that transforms $H_m$
into $V_{rs}^T$ had been thoroughly checked for excitation energies
but not for binding energies. It had a
bug in it that accounts for nearly -250 keV in
the $a$ term. Hence, when ${\cal S}$ is taken into account and
the bug is corrected, the fit becomes
(in {\em KeV}) $\varepsilon=-50$, $a=-59$, $b=235$. The numbers are
now pleasingly small and the principal uncertainty stems from the
parameters in ${\cal S}$.

Our mass results are shown in
Figs.~\ref{f:fig3}a,b. While there is both underbinding and overbinding
of the nuclei studied, the agreement is reasonably acceptable. It
becomes remarkable if we consider that --- in view of the smallness of
$H_{corr}$ --- it is practically parameter free.
For completeness, in Fig.~\ref{f:fig3}c we show
$\langle  H_{CM} \rangle$ for the same nuclei. Notice that for the nuclei
above mass 40 the center-of-mass contamination could be further corrected
by fine tuning $\beta_{CM}$. However, for our present purposes, we will
be content with the removal of much of the center-of-mass energy.

As a final example of the soundness of the interaction, we show in
Fig.~\ref{f:fig4} a number of low-lying states for $^{22}$Mg
calculated by direct diagonalization  in the
full $sd$-$pf$ space, compared to both a $sd$-shell calculation using
the USD interaction and to experiment \cite{r:ensdf}.  Generally, our
interaction agrees reasonably well with both experiment and the USD
interaction. The more refined treatment of $H_m$ will no doubt further
improve the agreement.  We also note that we have checked
the center-of-mass contamination for all of the excited states shown in
the first column of Fig.~\ref{f:fig4}, and it is as small as that
shown for the ground state in Fig.~2.

\section{Results}

\subsection{Comparison with experiment and other calculations}

There is limited experimental information about the highly unstable,
neutron-rich nuclei under consideration.  In many cases only the mass,
excitation energy of the first excited state, the $B(E2)$ to that state,
and the $\beta$-decay rate is known, and not even all of this
information is available in some cases.  From the
measured $B(E2)$, an estimate of the nuclear deformation parameter,
$\beta_2$, has been obtained via the usual relation
\begin{equation}
\beta_2 = 4 \pi \sqrt{B(E2; 0^+_{gs} \rightarrow 2^+_1)}/3 Z R_0^2 e
\end{equation}
with $R_0 = 1.2 A^{1/3}$ fm and $B(E2)$ given in $e^2$fm$^4$.

Much of the interest in the region stems from the unexpectedly large
values of the deduced $\beta_2$, results which suggest the onset of
deformation and have led to speculations about the vanishing of the $N
= 20$ and $N = 28$ shell gaps.  The lowering in energy of the 2$^+_1$
state supports this interpretation.  The most thoroughly studied case,
and the one which most convincingly demonstrates these phenomena, is
\mg32 with its extremely large $B(E2) = 454 \pm 78 \, e^2$fm$^4$ and
corresponding $\beta_2 = 0.513$ \cite{r:motobayashi}; however, a word of
caution is necessary when deciding on the basis of this
limited information that we are in the presence of well-deformed
rotors: for $^{22}$Mg, we would obtain $\beta_2 = 0.67$, even more
spectacular, and for $^{12}$C, $\beta_2 = 0.8$, well above the
superdeformed bands.

Most of the measured observables can be calculated within the SMMC
framework.  It is well known that in {\it deformed} nuclei the total
$B(E2)$ strength is almost saturated by the $0^+_{gs} \rightarrow
2_1^+$ transition (typically 80\% to 90\% of the strength lies in this
transition).  Thus the total strength calculated by SMMC should only
slightly overestimate the strength of the measured transition.  In
Table \ref{t:tab1} the SMMC computed values of $B(E2, total)$ are
compared both to the experimental $B(E2; 0^+_{gs} \rightarrow 2^+_1)$
values and to the values found in various truncated shell-model
calculations.  Reasonable agreement with experimental data across the
space is obtained when one chooses effective charges of $e_p=1.5$ and
$e_n=0.5$. We also indicate in the right column of Table \ref{t:tab1}
the USD values for the $B(E2,0_{gs}^+ \rightarrow 2_1^+)$ (with
effective charges of $e_p=1.5$ and $e_n=0.5$) for the $sd$-shell
nuclei. Note that the $sd$-shell results are much lower for $^{30}$Ne
and $^{32}$Mg than is seen experimentally. All of the theoretical
calculations require excitations to the $pf$-shell before reasonable
values can be obtained.  We note a general agreement among all
calculations of the $B(E2)$ for $^{46}$Ar, although they are typically
larger than experimental data would suggest. We also note a somewhat
lower value of the $B(E2)$ in this calculation as compared to
experiment and other theoretical calculations in the case of $^{42}$S.
Shown in Table \ref{t:tab2} are effective charges from other
calculations.

Table \ref{t:tab3} gives selected occupation numbers for the nuclei
considered.  We first note a difficulty in extrapolating some of the
occupations where the number of particles is nearly zero.  This leads
to a systematic error bar that we estimate at $\pm 0.2$ for all
occupations shown, while the statistical error bar is quoted in the
table. The extrapolations for occupation numbers were principally
linear. Table \ref{t:tab3} shows that $^{22}$Mg remains as an almost
pure $sd$-shell nucleus, as expected.  We also see that the protons in
$^{30}$Ne, \mg32, and $^{42}$Si are almost entirely confined to the
$sd$ shell.  This latter is a pleasing result in at least two regards.
First, it shows that the interaction does not mix the two shells to an
unrealistically large extent.  Second, if spurious center-of-mass contamination
were a severe problem, we would expect to see a larger proton
$f_{7/2}$ population for these nuclei due to the $0d_{5/2}$-$0f_{7/2}$
``transition'' mediated by the center-of-mass creation operator.  The fact that
there is little proton $f_{7/2}$ occupation for these nuclei confirms
that the center-of-mass contamination is under reasonable control.

An interesting feature of Table \ref{t:tab3} lies in the neutron
occupations of the $N = 20$ nuclei ($^{30}$Ne and \mg32) and the $N =
28$ nuclei ($^{42}$Si, $^{44}$S, and $^{46}$Ar).  The neutron
occupations of the two $N = 20$ nuclei are quite similar, confirming
the finding of Fukunishi {\it et al.} \cite{r:fukunishi} and Poves and
Retamosa \cite{r:poves1} that the $N= 20$ shell gap is modified.  In
fact, the neutron $f_{7/2}$ orbital contains approximately two
particles before the $N=20$ closure, thus behaving like an intruder
single-particle state.  Furthermore, we see that 2p-2h excitations
dominate although higher excitations also play some role.  We also see
that the neutrons occupying the $pf$-shell in $N=20$ systems are
principally confined to the $f_{7/2}$ sub-shell.

The conclusions that follow from looking at nuclei with $N > 20$,
particularly those with $N = 28$, are that the $N = 20$ shell is nearly
completely closed at this point, and that the $N=28$ closure shell is
reasonably robust, although approximately one neutron occupies the upper
part of the $pf$ shell. Coupling of the protons with the low-lying
neutron excitations probably accounts for the relatively large
$B(E2)$, without the need of invoking rotational behavior.

In Table \ref{t:tab4} we show the SMMC total Gamow-Teller (GT$^-$)
strength.  We compare our results to those of previous truncated
calculations, where available.  In all cases, our results are slightly
smaller than, but in good accord with, other calculations.  Since we
do not calculate the strength function, we do not compute
$\beta$-decay lifetimes.

\subsection{Pairing properties}

For a given angular momentum $J$, isospin $T$, and
parity $\pi$, we define the pair operators as
\begin{equation}
A^\dagger_{JM,TT_z\pi} (a b) = \frac{(-1)^{l_a}}{\sqrt{1+\delta_{ab}}}
[ a_{j_a}^{\dagger}\times a_{j_b}^{\dagger} ]^{JM,TT_z}\;,
\end{equation}
where the parity is given by $(-1)^{l_a+l_b}$.
These operators are boson-like in the sense that
they satisfy the expected commutation relations in the limit
where the number of valence nucleons is small compared with  the total
number  of single-particle states in the shell.  In the SMMC we
compute the pair matrix  in the ground state as
\begin{equation}
M_{JTT_z\pi}(ab,cd) = \sum_M\langle A^{\dagger}_{JM,TT_z\pi}(a
b)A_{JM,TT_z\pi}(
c d)\rangle\;,
\end{equation}
which
is a hermitian and positive-definite matrix in the space of
ordered orbital pairs $(a b)$ (with $a\leq b$).
The total number of pairs is given by
\begin{equation}
P_{JTT_z\pi}=\sum_{abcd} M_{JTT_z\pi}(ab,cd)\;.
\end{equation}
The pair matrix can be diagonalized to find the eigenbosons
$B^{\dagger}_{\alpha JTT_z\pi}$ as
\begin{equation}
B^\dagger_{\alpha JMTT_z\pi}=\sum_{a b} \psi_{\alpha JT\pi}(a b)
A^{\dagger}_{JMTT_z\pi}(a b)\;,
\end{equation}
where $\alpha=1,2,\cdots$ labels the
various bosons with the same angular momentum, isospin, and parity. The
$\psi_{\alpha JT\pi}$ are the eigenvectors of the diagonalization, i.e.
the wavefunctions of the boson, and satisfy the
relation
\begin{equation}
\sum_{j_aj_b}
\psi^*_{\alpha JTT_z\pi}\psi_{\mu JTT_z\pi} =\delta_{\alpha\mu}\;.
\end{equation}
These eigenbosons satisfy
\begin{eqnarray}\label{9}
\sum_M \langle B_{\alpha JM,TT_z\pi}^\dagger  B_{\gamma JM,TT_z\pi}\rangle =
n_{\alpha JTT_z\pi} \delta_{\alpha \gamma}\;,
\end{eqnarray}
where the positive eigenvalues $n_{\alpha JTT_z\pi}$ are  the number of
$JTT_z\pi$-pairs of type $\alpha$.

We first show the number of pairs $P_{JTT_z\pi}$ in the
$J=0$, $T=1$ positive-parity pairing channels. This quantity can be
interpreted as the total strength for the pair transfer of particles
of the given quantum numbers.
Shown in Fig.~\ref{f:fig5} are our results in the proton-neutron
(top), proton-proton (middle), and neutron-neutron (bottom) channels as
a function of the nucleus, $A$. Notice that only in the $N$=$Z$ nucleus
$^{36}$Ar do the proton-neutron pairs play a significant role, as
has been discussed in \cite{r:langanke}. Generally, one also sees
an increase in the proton-proton pairs as $A$ is increased. Notice
also that a fair amount of increase occurs in the sulfur and argonne
isotope chains as one adds neutrons. This is not the case in the
two Mg isotopes calculated, in which we see a significant increase
in the neutron-neutron correlations, but very little change in
the proton-proton sector. This holds for both the Ne and Mg chains
\cite{dean98}.  For the heavier isotopes in the
region, in general, the $J=0$
neutron-neutron pairs are not significantly enhanced for the
nuclei that we have calculated here. Since there are many
more particles and hence more pairing, one expects enhancements
to occur in higher $J$ pairs since the total number of pairs is
a conserved quantity for a given number of like nucleons.
We also calculated the pairing in the same channels, but with negative
parity ($\pi = (-1)^{l_a + l_b} = (-1)^{l_c + l_d}$),
and find it to be rather small in most cases.

Further insight into the pairing comes by considering diagonal
elements of the pair matrix before and after diagonalization.
The presence of a pair condensate in a correlated ground state will
be signaled by the largest eigenvalue for a given $J$ being
much greater than any of the others.
Shown in Fig.~\ref{f:fig6} are the diagonal matrix elements
of the $J=0$ pair matrix for $^{40,44}$S
before (left panel) and after (right panel) diagonalization.
We see from the left panel that adding four neutrons to the system
increases the $f_{5/2}$-shell neutron matrix elements,
while rearranging the
$sd$-shell elements slightly.
>From the occupation numbers we know that the neutrons are filling
$pf$-shell orbitals, and therefore we expect little movement
in the $sd$ shell.
The proton matrix elements are slightly affected by
the addition of the neutrons, although there is
some movement of protons out of the $f_{7/2}$.

The largest eigenvalue of the neutron-neutron pair matrix, as
shown in the right panel in Fig.~\ref{f:fig6},
is about 1.5 times that of the next largest eigenvalue.
However, the remaining eigenvalues are significant.
Thus it is unlikely that there
exists  a pure pair condensate in the neutrons.
As a further check on this conclusion, we have diagonalized the
3$\times$3 pairing matrix resulting from only the $sd$-shell
neutrons in these two nuclei.  We find that the three eigenvalues are
all of similar size and significantly smaller than the largest eigenvalue
from the full $sd$-$pf$ diagonalization.  Thus, what
neutron pair condensate does
exist is a phenomenon which involves the entire model space, not just
the $sd$ shell.
In the proton sector we see a similar level of
pair condensation. Since the protons occupy
mainly the $sd$-shell, only three eigenvalues are large enough to be
represented in the figure.

\subsection{Discussion}

The aim of a nuclear structure calculation is to compare with, or
predict, experimental results. In the present case, the comparison with
other calculations is at least of equal interest. The reason comes
from the problems created by truncations, in particular the (0+2)\hw
``catastrophe'' \cite{r:wbmb}, discovered long ago in a 0\hw
context~\cite{pasquini}. Calling $f$ the $f_{7/2}$ shell and $r$ its
$pf$ partners generically, an $f^n$ calculation can produce very
sensible results; $f^n+f^{n-1}r$ improves them considerably, but
$f^n+f^{n-1}r+f^{n-2}r^2$ is invariably disastrous, because the $f^n$
configuration is strongly pushed down by pairing with $f^{n-2}r^2$,
while $f^{n-1}r$ does not benefit from a similar push from
$f^{n-3}r^2$. The remarkable thing is that when this last
configuration is included, the results are not too different from the
original $f^n+f^{n-1}r$. If the space is expanded, there is an
attenuated 4\hw catastrophy. The process continues in increasingly
attenuated form until the exact $(fr)^n$ space is reached. This is a
general problem with truncations, and for nuclei such as $^{32}$Mg
with a $(pf)^2(sd)^{n-2}$ ground state, where the calculations demand
also the presence of $(sd)^n$ states , the adopted solution has been
simply to ignore the mixing between the two configurations.  It works
very well. But is it true? Could it not be possible that higher
excitations play an important role? To everybody's relief we can say
that the present calculations confirm the basic validity of previous
work. In all the cases we analyzed, the ``dressing'' process, whereby
a dominant configuration becomes the exact ground state, does not seem
to affect strongly its basic properties. Does it mean that exact
calculations are unnecessary? Not exactly. For one thing, they have no
parameters other than those of $H_m$ and therefore demonstrate the
validity of the monopole corrected $G$-matrices. And then they
go --- for the first time in a shell-model context --- to the heart of the
problem
of cross-shell correlations.  At present, we know little about these
problems,
except that they hide so well that they are difficult to detect.
Still, they can be seen
through effects (such as quenching of Gamow--Teller
strength) that tell us that they are important. The available
evidence points to a much reduced discontinuity at the Fermi level
with respect to the naive shell model \cite{PA84,BE93,CPZ95}. In
Table~\ref{t:tab3} we find nearly normal occupancies for high $T$, but
strong effects for $T=0$ ground states, in particular for $^{44}$Ti --- a
truly interesting case. A conventional $(pf)^4$ calculation yields a
$BE2$ of 514.7 $e^2f^4$ (virtually identical to that of KB3). In
Table~\ref{t:tab1} the result is at least 20\% larger. This is a good
example of the hiding talents of the correlations. No doubt this
nucleus is a {\it bona fide} member of the $pf$ space, and the
correlation effects can be drowned in the experimental error, but it
is not always the case. The region is plagued with BE2 transitions which are
 systematically too large for the 0\hw calculations to explain, particularly
for the Ca isotopes, which should be the simplest nuclei, but are the
most complicated.
In $^{44}$Ti we have a first example of what
a complete calculation could do.

Binding energies are no doubt one of the best ways to shed light
on the matter. In Section III.C we mentioned the
cross-shell gap around $^{40}$Ca, which should be increased by about
2 MeV, which means that the correlation energy should be much larger.
And since we know now that we can trust SMMC with a good interaction
to within 1 MeV, probably it will not take long before we know more
about this supposedly closed nucleus that is not so closed.

\section{Conclusion}

This paper was meant as a feasibility study of SMMC calculations in
multi-\hw spaces. Two general issues had to be tackled: translational
invariance and the definition of an interaction. Concerning the first,
it was shown that the trouble caused by center-of-mass excitations can
be successfully mitigated by a judicious application of ideas in
\cite{r:whitehead}, and a possible variational approach to the problem
was suggested. The interaction chosen was a $G$-matrix derived from a
modern potential, renormalized according to state-of-the-art
techniques, and monopole-corrected for the bad saturation properties
of the existing $NN$ potentials. The only parameters entering the
calculations are the six ``universal'' constants specifying the
monopole Hamiltonian, which was shown to explain quite naturally the
shell formation properties of high isospin nuclei in the $A$=30-50
region.

The feasibility test was passed satisfactorily. Binding energies,
$B(E2)$ rates, and Gamow--Teller strengths were obtained that are in
reasonable agreement with observations, and the possible origin of the
remaining discrepancies has been identified.

The calculations support the validity of previous work in the region,
and open the way to the study of the elusive deep correlations at the
origin of Gamow--Teller quenching. In particular it provides an
example, in $^{44}$Ti, of an extremely correlated system whose
behavior is quite similar to that of the uncorrelated one. The
possibility to obtain orbit occupancies should help in advancing the
study of the discontinuity at the Fermi surface --- one of the most
difficult problems in nuclear physics.

Interest was focused on neutron-rich nuclei around $N=$20,28 (a region
of current interest), where new data has become available and many
calculations have been performed. Having established the reliability
of our methods, other exotic, or not so exotic, studies can be
contemplated.

Most calculations presented here were
performed on the 512-node Paragon at the Oak Ridge Center for
Computational Science (CCS) and the T3E at the National Energy
Research Scientific Computing Center (NERSC). The $sd$-$pf$ model
space effectively used all of the available memory on the Paragon (32
Mbytes per node) and, hence, larger spaces were not feasible there.
With the advent of a new generation of massively parallel computers
that are much faster and have far more memory, much more ambitious
calculations are possible.

\acknowledgements
We acknowledge useful discussions with Petr Vogel and Dao-Chen Zheng.
M.T.R. gratefully acknowledges support from the Weingart
Foundation; he and S.E.K. were supported in part by
the U.S. National Science Foundation
under Grants PHY94-12818 and PHY94-20470.
Oak Ridge National Laboratory is managed by Lockheed Martin Energy
Research Corp. for the U.S. Department of Energy under contract number
DE-AC05-96OR22464.
K.L. acknowledges support from the Danish Research Council.
Grants of computational resources were provided by the Center
for Advanced Computational Research at Caltech and the
Center of Computational Science at ORNL, as well as
the National Energy Research Scientific Computing Center.
Part of this work was conducted at the Aspen Center for Physics.
DJD, KL, and AZ acknowledge support from a NATO travel grant
CRG.CRPG 973035.


\begin{figure}
\caption[Figure 1] {(a)  The calculated value of
$\langle  H_{CM} \rangle$ as a function of
$\beta_{CM}$ for \mg32.  Two different extrapolations
were performed as described in the text.  The center-of-mass contamination is
already significantly reduced at $\beta_{CM} = 1$.
(b)  The calculated total $B(E2, 0^+ \rightarrow 2^+)$ as a function
of $\beta_{CM}$.  (c)  The $sd$-shell and $f_{7/2}$ subshell occupations
as a function of $\beta_{CM}$.
}
\label{f:fig1}
\end{figure}

\begin{figure}
\caption[Figure 2] {(a).  The expectation of the Hamiltonian,
$\langle H\rangle$ for $^{22}$Mg as a function of the extrapolation
parameter $g$. Shown are standard shell-model results and SMMC
results.  (b).  The expectation of the center-of-mass Hamiltonian,
$\langle H_{CM}\rangle$ as a function of $g$. SMMC results
are shown for two types of extrapolation procedures, as discussed
in the text, and are compared to standard shell-model results.
}
\label{f:fig2}
\end{figure}

\begin{figure}
\caption[Figure 3] { Monopole shell
effects in the binding energies of $T=4$ nuclei.}
\label{f:shell}
\end{figure}

\begin{figure}
\caption[Figure 4] {
(a).  The binding energy relative to the $^{16}$O core
for various nuclei in this study.  (b).  The difference
between experiment and theory for these nuclei.  (c).  The
expectation of the center-of-mass Hamiltonian for
the nuclei calculated in this study.
}
\label{f:fig3}
\end{figure}

\begin{figure}
\caption {Theoretical and experimental level spectra
for $^{22}$Mg are compared.  The left spectrum is
obtained from the Hamiltonian described in
the text.  USD is the Wildenthal $sd$-shell interaction
used in a $sd$-shell calculation for comparison.
}
\label{f:fig4}
\end{figure}

\begin{figure}
\caption { The number of pairs
present in the SMMC calculations for the
$J^{\pi}T = 0^+1$:  (a) the $T_z=0$ (pn) channel;
(b) the $T_z=1$ (pp) channel; (c) the
$T_z=-1$ (nn) channel.
}
\label{f:fig5}
\end{figure}

\begin{figure}
\caption {Left panel: the diagonal elements of the $J=0$ pp and nn
pair matrix before diagonalization. Note that proton pairing does not play
a significant role in the $pf$-shell.
Right panel: the eigenvalues of the pair
matrix shown in decreasing size. After diagonalization the protons have
only three non-zero eigenvalues.}
\label{f:fig6}
\end{figure}

\onecolumn

\begin{table}
\caption{Single-particle energies used in this study compared
to the two sets in refs.\protect\cite{dz98}.}
  \begin{center}
    \leavevmode
    \begin{tabular}{|r|r|r|r|}
\hline
 j p &  SPE old& SPE new& SPE calc.\\
\hline
  d$_{5/2}$ 3 &  16.679 &  15.193 &  15.129 \\
  s$_{1/2}$ 3 &  12.454 &  12.719 &  12.629 \\
  d$_{3/2}$ 3 &  10.404 &  10.543 &  10.629 \\
  f$_{7/2}$ 3 &   9.022 &   8.324 &   8.629 \\
  p$_{3/2}$ 2 &   6.381 &   6.133 &   5.595 \\
  p$_{1/2}$ 2 &   1.336 &   0.722 &   0.784 \\
  f$_{5/2}$ 2 &   0.000 &   0.000 &   0.000 \\
\hline
    \end{tabular}
    \label{t:sp}
 \end{center}
\end{table}

\begin{table}
\caption{The computed and measured values of $B(E2)$ for
the nuclei in this study using $e_p=1.5$ and $e_n=0.5$.
}
\begin{tabular}{ccccc}
 & $B(E2; 0^+_{gs} \rightarrow 2^+_1)_{Expt}$ & $B(E2, total)_{SMMC}$ &
  $B(E2; 0^+_{gs} \rightarrow 2^+_1)$ &
$B(E2,0^+_{gs} \rightarrow 2^+_1)_{USD}$ \\
\tableline
 & & & & \\
 $^{22}$Mg & $458 \pm 183$ & $334 \pm 27 $
    & & 314.5 \\
 $^{30}$Ne & & $303 \pm 32$
    & 342\cite{r:fukunishi},171\cite{r:poves2} & 143.2 \\
 $^{32}$Mg & $454 \pm 78$ \cite{r:motobayashi} & $494 \pm 44 $
   & 448
\cite{r:fukunishi},205\cite{r:poves2} & 177.1 \\
 $^{36}$Ar & $296.56 \pm 28.3$\cite{r:ensdf} & $174 \pm 48$
    & & 272.8 \\
 $^{40}$S & $334 \pm 36$ \cite{r:brown1} & $270 \pm 66$
    & 398\cite{r:brown2},390\cite{r:retamosa} & \\
 $^{42}$S & $397 \pm 63$ \cite{r:brown1} & $194 \pm 64$
   & 372\cite{r:brown2},465\cite{r:retamosa} & \\
 $^{42}$Si &  &  $445 \pm 62$
    & 260\cite{r:retamosa} & \\
 $^{44}$S & $314 \pm 88$ \cite{r:brown2} & $274 \pm 68$
    & 271\cite{r:brown2},390\cite{r:retamosa} & \\
 $^{44}$Ti & $610 \pm 150$ \cite{r:raman} & $692 \pm 63$
    & & \\
 $^{46}$Ar & $196 \pm 39$ \cite{r:brown1} & $369 \pm 77 $
    & 460\cite{r:brown1},455\cite{r:retamosa} & \\
\end{tabular}
\label{t:tab1}
\end{table}


\begin{table}
\caption{The effective charges $e_p$ and $e_n$ used in the
various truncated shell-model calculations for the nuclei in this study.}
\begin{tabular}{ccc}
Reference & $e_p$ & $e_n$ \\
\tableline
\cite{r:brown1} & 1.6 & 0.9 \\
\cite{r:brown2} & 1.35 & 0.65 \\
\cite{r:poves1,r:retamosa} & 1.5 & 0.5 \\
\cite{r:fukunishi} & 1.3 & 0.5 \\
\end{tabular}
\label{t:tab2}
\end{table}


\begin{table}
\caption{The calculated SMMC neutron ($n$) and
proton ($p$) occupation numbers for the $sd$ shell, the
$0f_{7/2}$ sub-shell, and the remaining orbitals of the
$pf$ shell.  The statistical errors are given for linear
extrapolations. A systematic error of $\pm 0.2$ should also
be included.}
\begin{tabular}{c|c|ccc|ccc}
 & $N,Z$ & $n$-$sd$ & $n$-$f_{7/2}$ & $n$-$pf_{5/2}$ &
   $p$-$sd$ & $p$-$f_{7/2}$ & $p$-$pf_{5/2}$ \\
\tableline
$^{22}$Mg & 10,12 & $3.93 \pm 0.02$ & $0.1 \pm  0.02$ &
  $-0.05 \pm 0.01$ & $2.04 \pm 0.02$ & $0.00 \pm 0.01$ &
  $-0.05 \pm 0.01$ \\
$^{30}$Ne & 20,10 & $9.95 \pm 0.03$ & $2.32 \pm 0.03$ &
  $-0.26 \pm 0.02$ & $2.03 \pm 0.02$ & $-0.01 \pm 0.01$ &
  $-0.02 \pm 0.01$ \\
$^{32}$Mg & 20,12 & $9.84 \pm 0.03$ & $ 2.37 \pm 0.03$ &
  $-0.21 \pm 0.02$ & $3.99 \pm 0.03$ & $0.05 \pm 0.02$ &
  $-0.05 \pm 0.01$ \\
$^{36}$Ar & 18,18 & $9.07 \pm 0.03$ & $1.08 \pm 0.02$ &
  $-0.15 \pm 0.02$ & $9.07 \pm 0.03$ & $1.08 \pm 0.02$ &
  $-0.15 \pm 0.02$ \\
$^{40}$S & 24,16 & $11.00 \pm 0.03$ & $ 5.00 \pm 0.03 $ &
  $-0.01\pm 0.02$ & $7.57 \pm 0.04$ & $0.54 \pm 0.02$ &
  $-0.12 \pm 0.02$ \\
$^{42}$Si & 28,14 & $11.77 \pm 0.02$ & $7.34 \pm 0.02$ &
  $0.90 \pm 0.03$ & $5.79 \pm 0.03$ & $0.25 \pm 0.02$ &
  $-0.07 \pm 0.01$ \\
$^{42}$S & 26,16 & $11.41 \pm 0.02$ & $6.33 \pm 0.02$ &
  $0.25 \pm 0.03$ & $7.49 \pm 0.03$ & $0.58 \pm 0.02$ &
  $-0.09 \pm 0.02$ \\
$^{44}$S & 28,16 & $11.74 \pm 0.02$ & $7.18 \pm 0.02$ &
  $1.06 \pm 0.03$ & $7.54 \pm 0.03$ & $0.56 \pm 0.02$ &
  $-0.12 \pm 0.02$ \\
$^{44}$Ti & 22,22 & $10.42 \pm 0.03$ & $3.58 \pm 0.02$ &
  $0.00 \pm 0.02$ & $10.42 \pm 0.03$ & $3.58 \pm 0.02$ &
  $0.00 \pm 0.02$ \\
$^{46}$Ar & 28,18 & $11.64 \pm 0.02$ & $7.13 \pm 0.02$ &
  $1.23 \pm 0.03$ & $8.74 \pm 0.03$ & $1.34 \pm 0.02$ &
  $-0.08 \pm 0.02$ \\
\end{tabular}
\label{t:tab3}
\end{table}


\begin{table}
\caption{The calculated total Gamow-Teller strength, $GT^-$,
from this study.  The results of other studies, when
available, are presented for comparison.
}
\begin{tabular}{ccc}
 Nucleus & SMMC & Other \\
\tableline
 $^{22}$Mg & $0.578 \pm  0.06$  & \\
 $^{30}$Ne & $29.41 \pm 0.25$ & \\
 $^{32}$Mg & $24.00 \pm 0.34$ & \\
 $^{36}$Ar & $2.13 \pm  0.61$ & \\
 $^{40}$S  & $22.19 \pm 0.44$ & 22.87\cite{r:retamosa} \\
 $^{42}$S  & $28.13 \pm 0.42$ & 28.89\cite{r:retamosa} \\
 $^{42}$Si & $40.61 \pm 0.34$ & \\
 $^{44}$S  & $34.59 \pm 0.39$ & 34.93\cite{r:retamosa} \\
 $^{44}$Ti & $4.64 \pm  0.66$ & \\
 $^{46}$Ar & $29.07 \pm 0.44$ & 28.84\cite{r:retamosa} \\
\end{tabular}
\label{t:tab4}
\end{table}

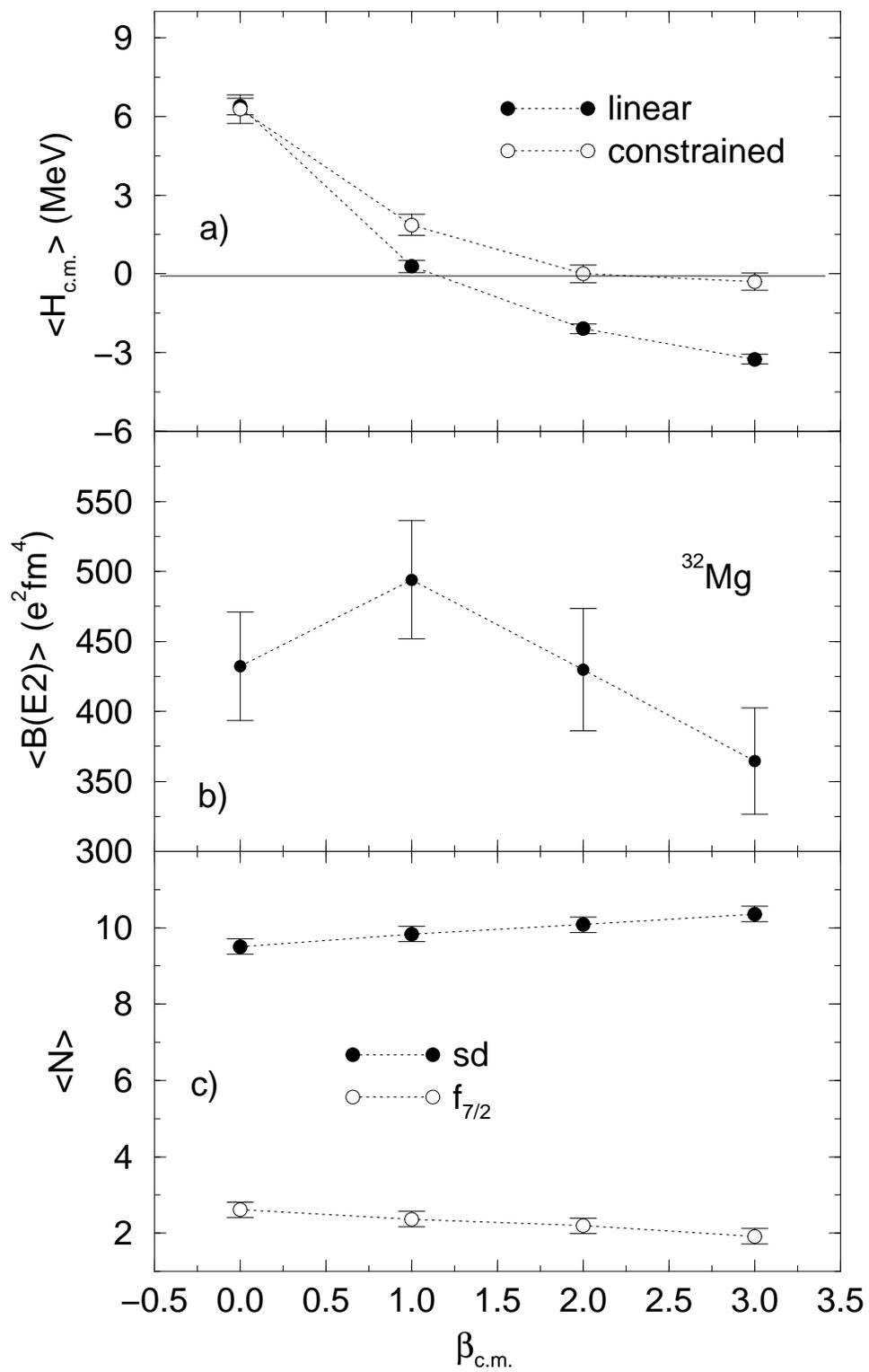
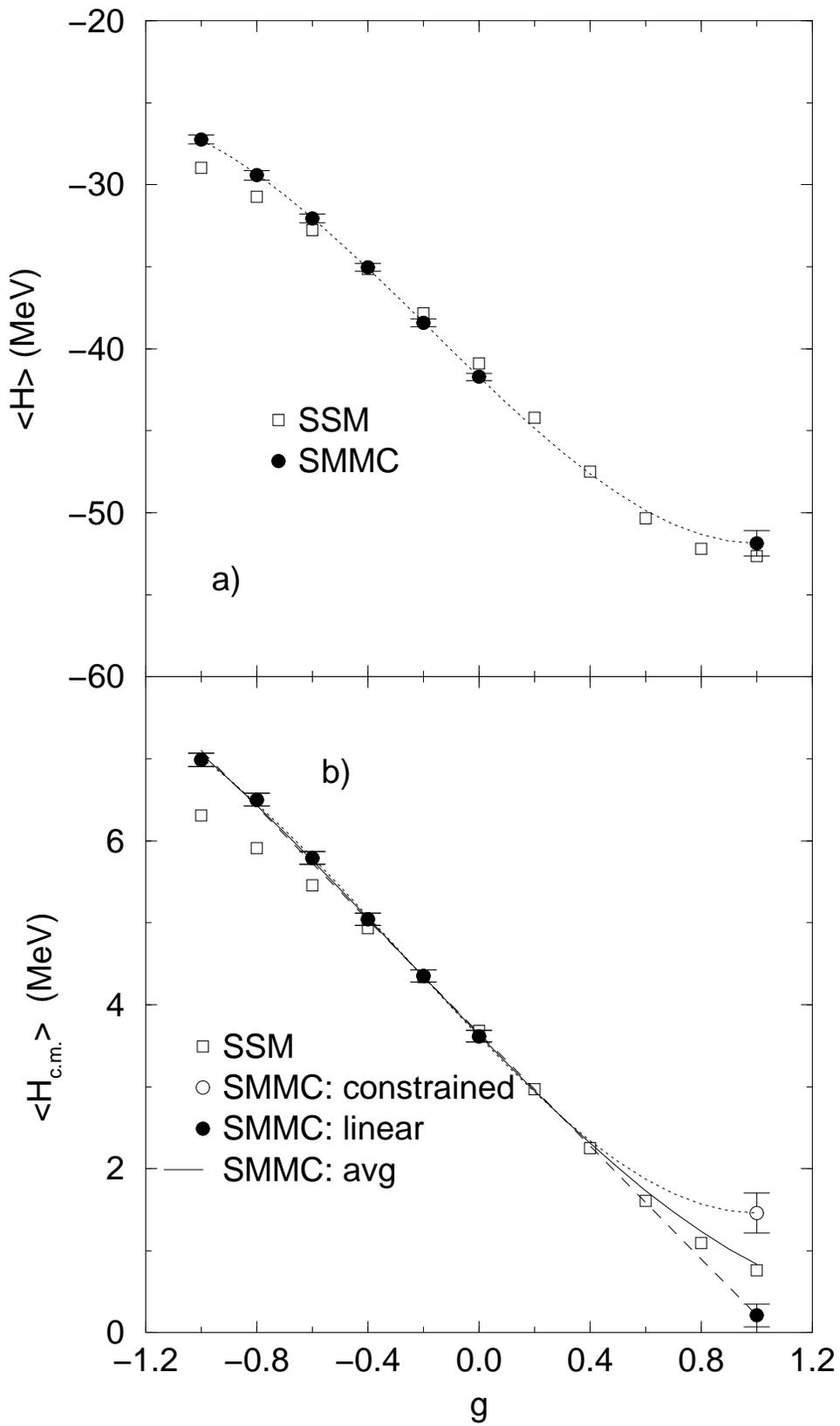
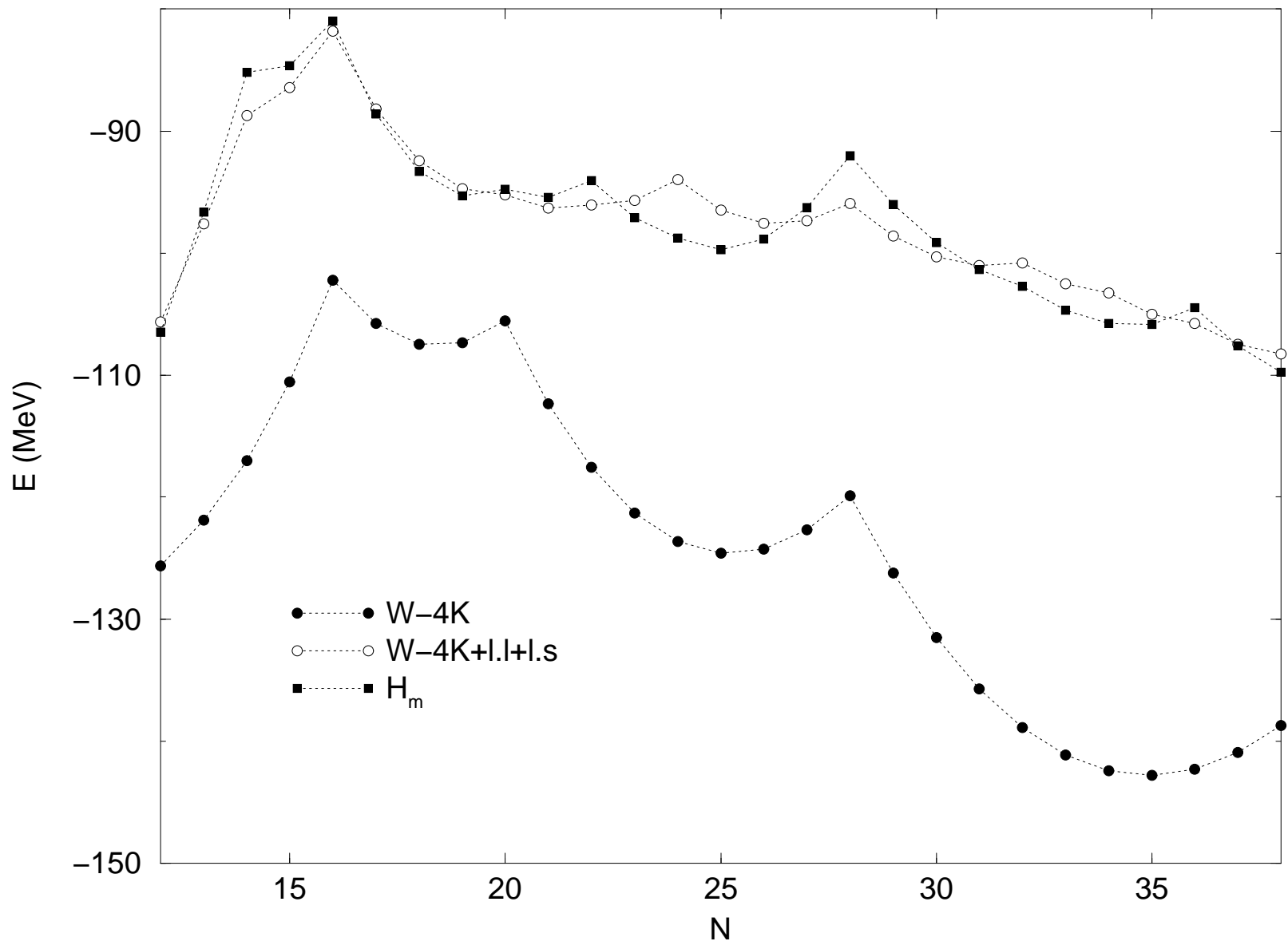
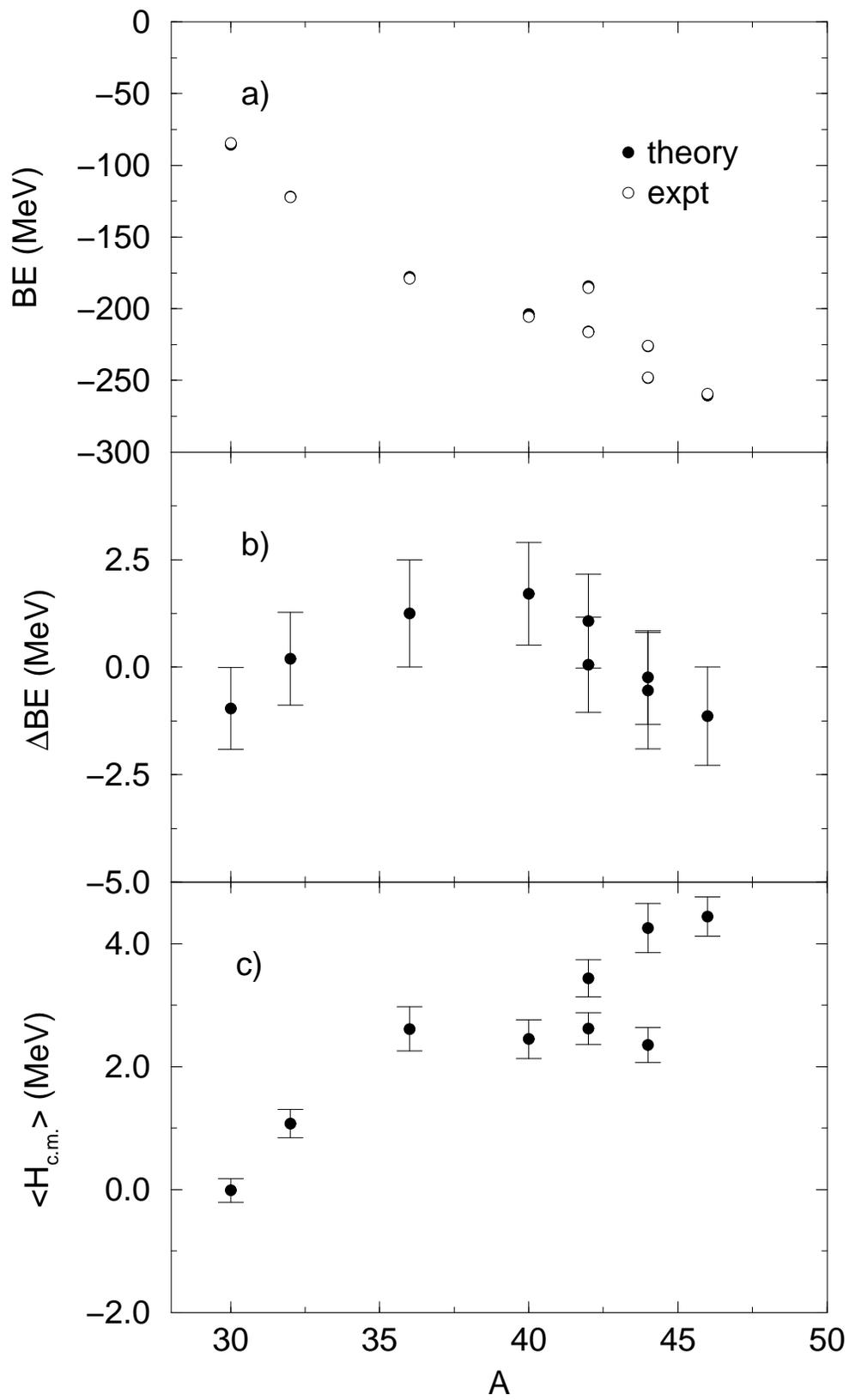
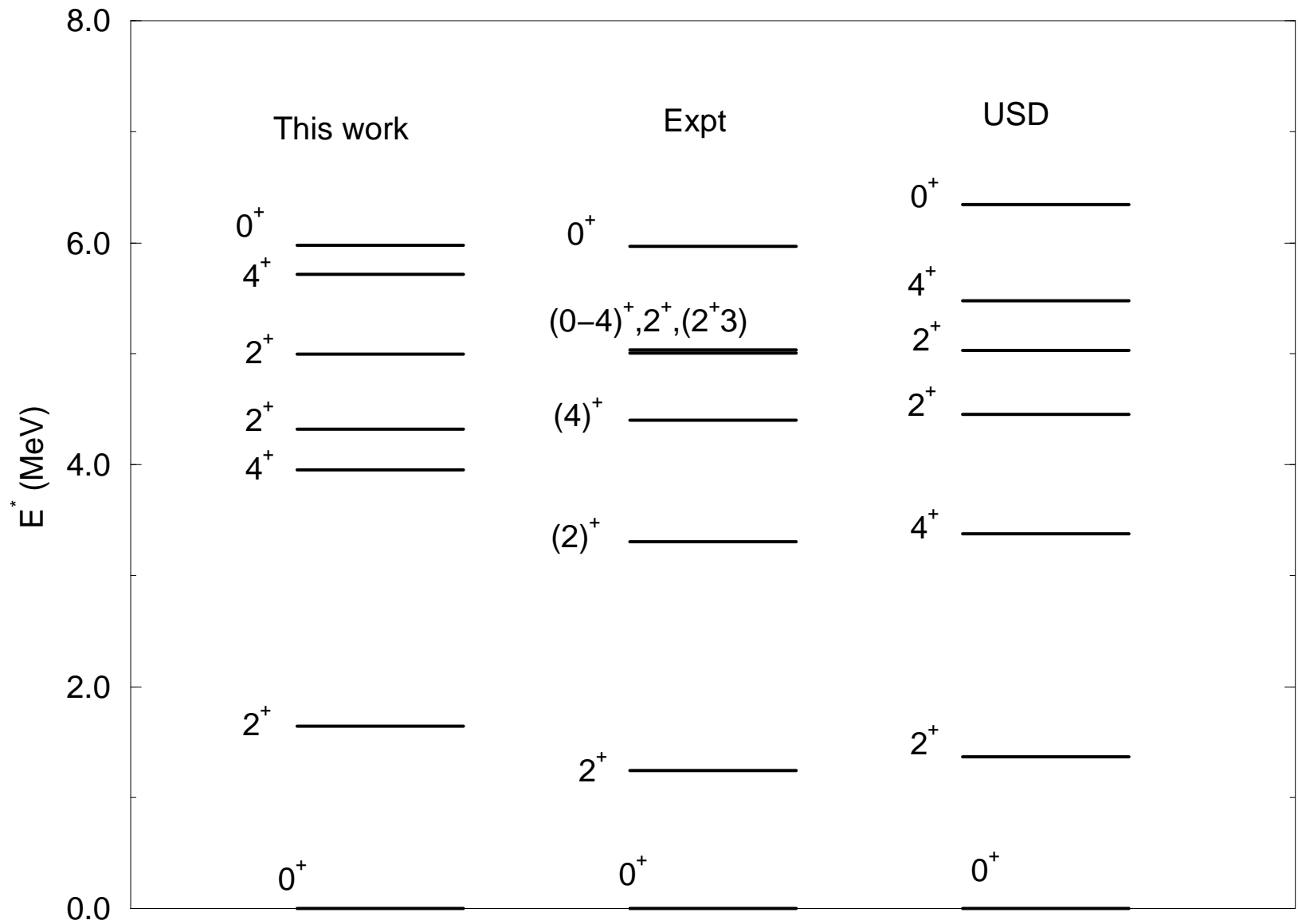
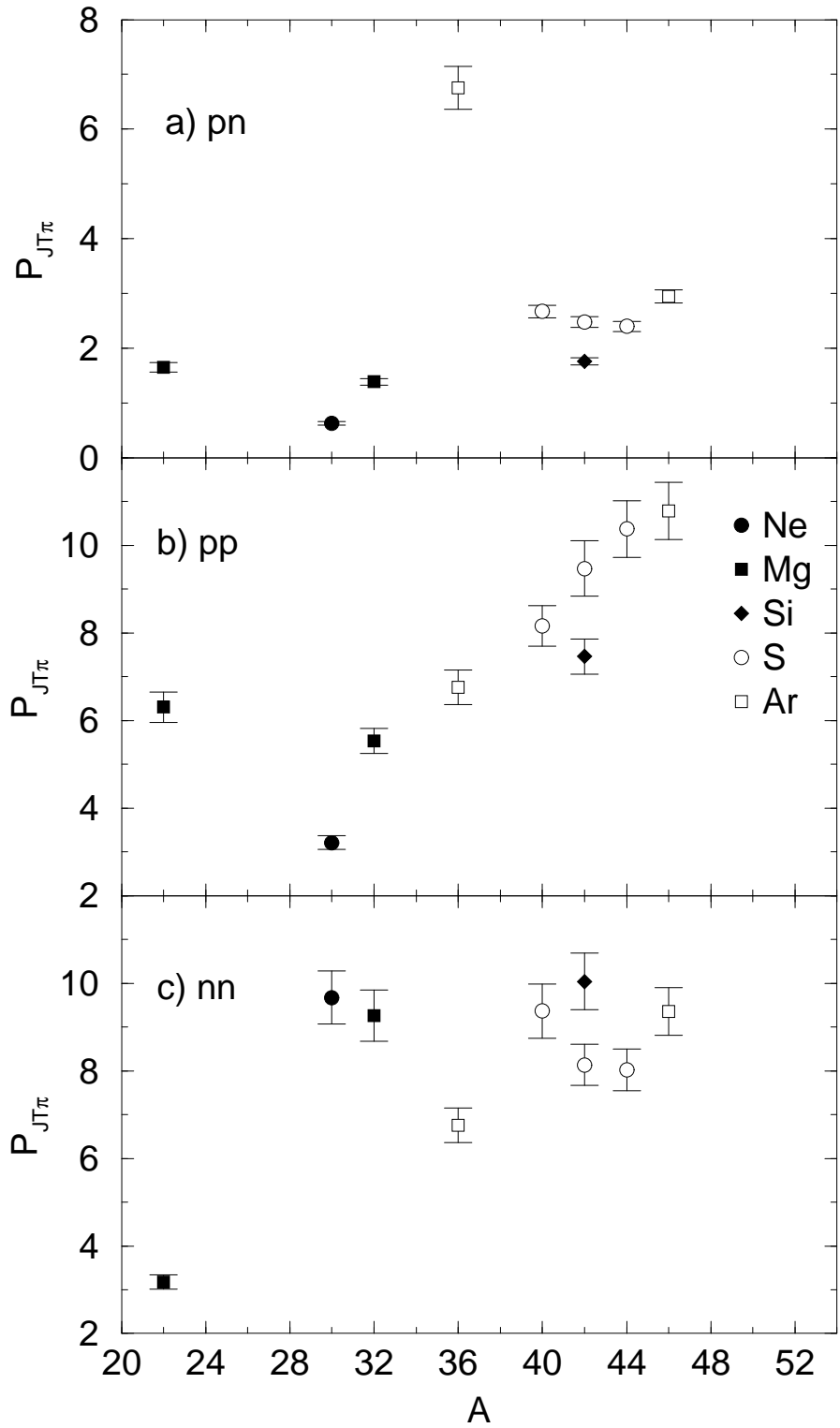
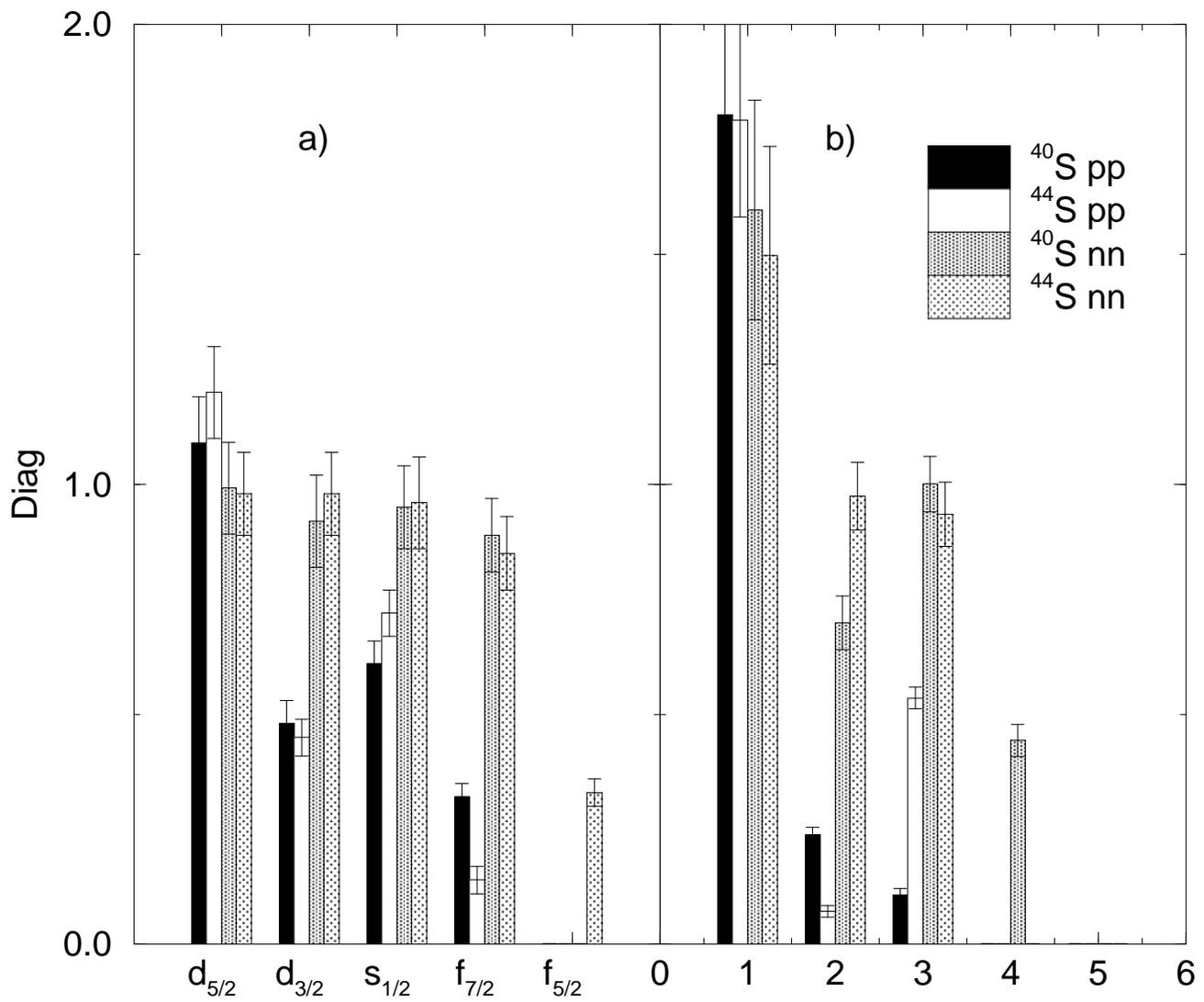
\end{document}